# Photogeneration and signatures of coherent phonons in time-resolved photoemission spectroscopy: First-principles time-dependent adiabatic GW approach

Yang-Hao Chan,[1,2] Zhenglu Li,[3,4,5] and Steven G. Louie[4,5]

[1]*Institute of Atomic and Molecular Sciences, Academia Sinica, Taipei 10617, Taiwan*

[2]*Physics Division, National Center of Theoretical Sciences, Taipei 10617, Taiwan*

[3]*Mork Family Department of Chemical Engineering and Materials Science, University of Southern California, Los Angeles, California 90089, USA*

[4]*Department of Physics, University of California at Berkeley, Berkeley, California, 94720, USA*

[5]*Materials Sciences Division, Lawrence Berkeley National Laboratory, Berkeley, California 94720, USA*

Coherent lattice dynamics can be observed in pump-probe time-resolved and angle-resolved photoemission spectroscopy (TR-ARPES) as a periodic modulation of intensity and energy of photoelectrons over probe time. We present an *ab initio* time-dependent GW approach including electron-phonon (e-ph) couplings to simulate the photogeneration of coherent phonons and their effects on the TR-ARPES of monolayer $MoS_2$. We demonstrate that state-resolved e-ph coupling strength can be obtained from analyzing coherent phonon modulations on TR-ARPES. Features of both the impulsive stimulated Raman scattering mechanism and the displacive excitation mechanism of coherent phonon generation are identified in our simulations. We clarify their origins and the coincident selection rule of coherent phonon generation and Raman scattering intensity. This method provides supports to analyze coherent phonon dynamics and e-ph couplings in TR-ARPES and enables quantitative engineering of band structure through coherent phonons.

# I. INTRODUCTION

Coherent lattice dynamics, or phonons, excited by light pulses can modulate electronic and lattice properties. Besides the fundamental interest in understanding the interplay between light-matter and electron-phonon (e-ph) couplings [1], coherent phonons serve as a tuning knob for modifying the optical [2, 3] and magnetic properties [4] of materials, ferroelectric transitions [5], or even driving matter into novel metastable states, with promising applications in quantum control realization [6–8] as well as sensors and transducers [9].

Coherent phonons have been studied mostly by optical pump-probe experiments. Pump light is used to generate coherent phonons and then their modulation of the optical response can be detected via transient reflectivity, absorption, or transmission [10, 11] measurements. Moreover, the presence of coherent phonons can induce exciton resonance peak shift in the spectrum, which is often considered as a spectral fingerprint of exciton-phonon coupling [12, 13]. However, to date, detailed information on state-resolved exciton-phonon or e-ph coupling strength is still out-of-reach from such optical measurements.

Time- and angle-resolved photoemission spectroscopy (TR-ARPES) is a well-established tool for studying electron excitation dynamics. The accessibility of electronic structure in the momentum space enables the detection of subtle changes in electronic structure. Coherent phonon-modulated TR-ARPES in principle can reveal state-resolved e-ph coupling strength through Fourier analysis of ARPES signals [14]. A direct determination of deformation potential has been reported by coherent lock-in measurement of TR x-ray diffraction and TR-ARPES [15]. Moreover, evidence of coherent phonon-mediated electronic structure control and coherent phonon-induced topological phase transition in $1T_d$ $WTe_2$ [16] and a metal-insulator transition in $Ta_2NiSe_5$ [17] were reported.

Theoretical description of coherent phonon has been addressed based on phenomenological models or formally in a density matrix formulation [18–22]. Two coherent-phonon generation mechanisms are often discussed in the literature for semiconductors. For pump frequencies in the transparent range, the impulsive stimulated Raman scattering (ISRS) generation of coherent phonon is believed to be the major cause, in which the generating force is closely related to the Raman tensor [19]. In the opaque frequency range, displacive excitation of coherent phonon (DECP) is understood to be due to the force exerted by the photo-excited carriers [18]. Typically, based on the driven harmonic oscillator model, a cosine oscillation is expected from the DECP mechanism while a sine oscillation is connected to the ISRS. However, debates remain as measured phase shifts between zero and $\pi$ have also been reported [23]. As real materials involve intricate couplings between many degrees of freedom, a first-principles calculation is necessary to fully address this problem.

Progress on understanding coherent phonon from first principles has been encouraging. Both coherent-phonon generation mechanisms discussed above have been explored at the level of time-dependent density functional theory (TDDFT) calculations, for Si [24] and Sb [25]. For monolayer $MoS_2$, where excitonic effects are significant, coherent phonon induced modulations on absorption spectrum was simulated in a static calculation by imposing displacements of the corresponding phonon mode in the atomic structure [26]. Absorption spectrum calculations with varying magnitudes of displacement were performed to compare with experiments. However, the method cannot address the interplay between electron and phonon dynamics, and it was by design not suitable to investigate the generation mechanism. Therefore, a time-dependent first-principles simulation of both coherent phonon generation and modifications in TR-ARPES for real materials that includes both electron–hole and e-ph interactions remains a major challenge.

Here, we extend our first-principles time-dependent adiabatic GW approach [28] (which includes electron-electron and electron-hole interactions) to incorporate e-ph couplings and compute the phonon dynamics and TR-ARPES spectrum in a pump-probe setup for a monolayer $MoS_2$. The approach addresses both the generation of coherent phonons and their effects on TR-ARPES. We demonstrate that detailed information on e-ph coupling strength can be obtained by analyzing the state-dependent modulations on the TR-ARPES intensity induced by coherent phonons. We show that the two coherent-phonon generation mechanisms arise from the same driving term but manifest themselves in two different frequency ranges, which agree with scenarios from previous theoretical and calculational results [20, 24]. We emphasize that the transition between the two mechanisms is not an abrupt one. We further clarify the relation between the selection rule of the coherent phonon generation and that of the Raman scattering and show that, although enhanced intensity of both is seen at energy in resonant with excitons in our calculations, the expressions derived from perturbation theory clearly indicate their differences in behaviors.

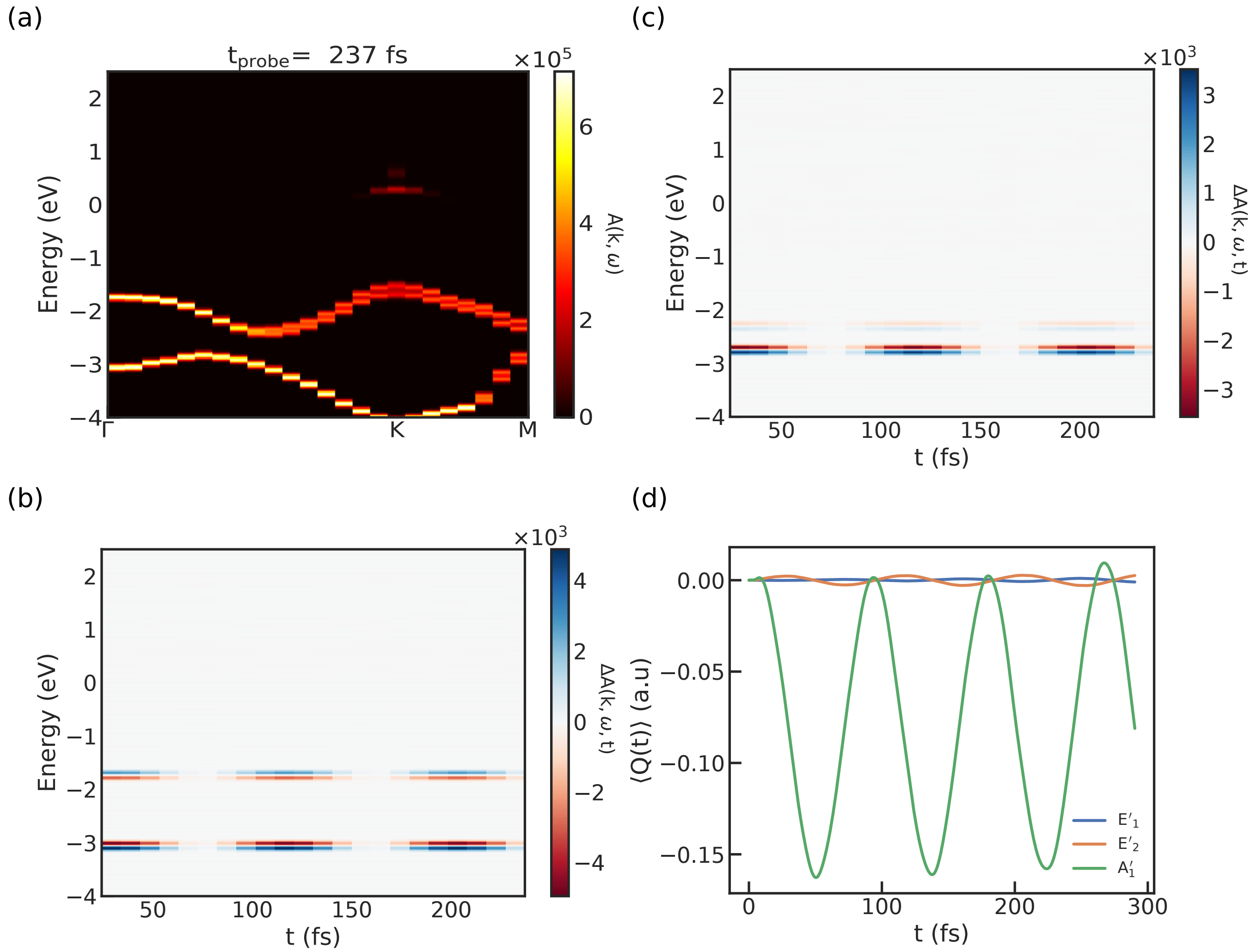

FIG. 1. (Color online) (a) Snapshot of computed spectral function $A(\boldsymbol{k},\omega,t)$, which reflects the TR-ARPES intensity, of monolayer $MoS_2$ at a probe time of 237 fs with a pump frequency of 1.7 eV of 10 fs duration and a maximum light field amplitude of $10^{-3}$ a.u. Momentum slices of spectral intensity difference $\Delta A(\boldsymbol{k},\omega,t)$ from (a) at (b) the $\boldsymbol{k}=\Gamma$ and (c) the $\boldsymbol{k}=M$ point. (d) Coherent phonon dynamics (i.e., phonon displacement $\langle Q_\nu(t)\rangle$) of the $A_1^{'}$ phonon mode and the doubly degenerate $E^{'}$ phonon mode from the same calculation.

## II. METHOD

We simulate the TR-ARPES spectrum for a monolayer $MoS_2$ with a pump light pulse having a time profile in the form of a truncated half-period sine squared function with a specific central frequency $\omega_c$, amplitude $E_{max}$, and a duration of 10 fs [27]. Immediately after the pump, we probe the system's electron spectral function every 10 fs with a probe duration of 50 fs to mimic the TR-ARPES spectrum. In our TD-aGW method [28], we propagate in time the density matrix in the Bloch band basis $\rho_{nm\boldsymbol{k}}$ (treating the position operator of the electron fully within the Berry connection formalism) and the phonon displacement $\langle Q_\nu(t)\rangle$ including electron-electron and electron-hole (e-h) (within the GW self-energy approximation) and e-ph interactions according to

$$i\hbar\frac{\partial\rho_{nm\boldsymbol{k}}}{\partial t} = [H(t),\rho(t)]_{nm\boldsymbol{k}}, \qquad (1)$$

with

$$H(t) = H^{QP} + \delta V^{H}(t) + \delta\Sigma^{COHSEX} + H^{e-ph}(t) + U^{ext}(t), \quad (2)$$

where $H^{QP}$ is the equilibrium GW quasiparticle Hamiltonian, $\delta V^{H}$ and $\delta\Sigma^{COHSEX}$ are the changes in the Hartree energy and the GW self-energy in the static COHSEX approximation from those at equilibrium, respectively, $U^{ext}$ is the coupling to the external field (i.e., the pump light pulse), and $H^{e-ph}(t)$ is the e-ph coupling term in the classical phonon approximation which reads

$$H^{e-ph}_{nm\boldsymbol{k}}(t) = \sum_{\nu}\tilde{g}_{nm\nu}(\boldsymbol{k},\boldsymbol{q}=0)\langle Q_\nu(t)\rangle, \quad (3)$$

with $\tilde{g}_{nm\nu\boldsymbol{k}} = \sqrt{2}N_p^{-\frac{1}{2}}g_{nm\nu}(\boldsymbol{k},\boldsymbol{q}=0)$ is the e-ph coupling matrix element scaled by the supercell size $N_p$ for a phonon with branch index $\nu$ and wavevector $\boldsymbol{q}=0$ mediating interactions between electron states $(n,\boldsymbol{k})$ and $(m,\boldsymbol{k}+\boldsymbol{q})$, and the displacement operator $Q_\nu = (b_{0\nu} + b^{\dagger}_{0\nu})/\sqrt{2}$ is defined from the phonon creation (annihilation) operator $b^{\dagger}_{0\nu}(b_{0\nu})$ with wavevector $\boldsymbol{q}=0$. Here, we focus on the zone-center phonons which are the most relevant modes for optical light induced coherent-phonon phenomena [18-20]. The equation of motion (EOM) for the expectation value of the displacement $\langle Q_\nu\rangle$ reads

$$\left(\frac{\partial^2}{\partial t^2} + \omega_{0\nu}^2\right)\langle Q_\nu(t)\rangle = -\frac{\omega_{0\nu}}{\hbar}\sum_{nm\boldsymbol{k}}\tilde{g}_{nm\nu\boldsymbol{k}}\rho_{mn\boldsymbol{k}}(t), \quad (4)$$

where $\omega_{0\nu}$ is the phonon frequency of mode $\nu$ and $\boldsymbol{q}=0$. Equation 4 is the EOM of a driven harmonic oscillator. The formal solution of $\langle Q_\nu\rangle$ in the frequency domain reads

$$\langle Q_\nu(\omega)\rangle = \frac{-\omega_{0\nu}}{\hbar(-\omega^2 + \omega_{0\nu}^2)}\sum_{nm\nu\boldsymbol{k}}\tilde{g}_{nm\nu\boldsymbol{k}}\rho_{mn\boldsymbol{k}}(\omega). \quad (5)$$

Equation 1-4 are coupled equations of the interacting single-electron density matrix and the expectation value of the phonon displacement. The details of the derivation, numerical time propagation, and the TR-ARPES calculations are given in the SM [27].

## III. RESULTS

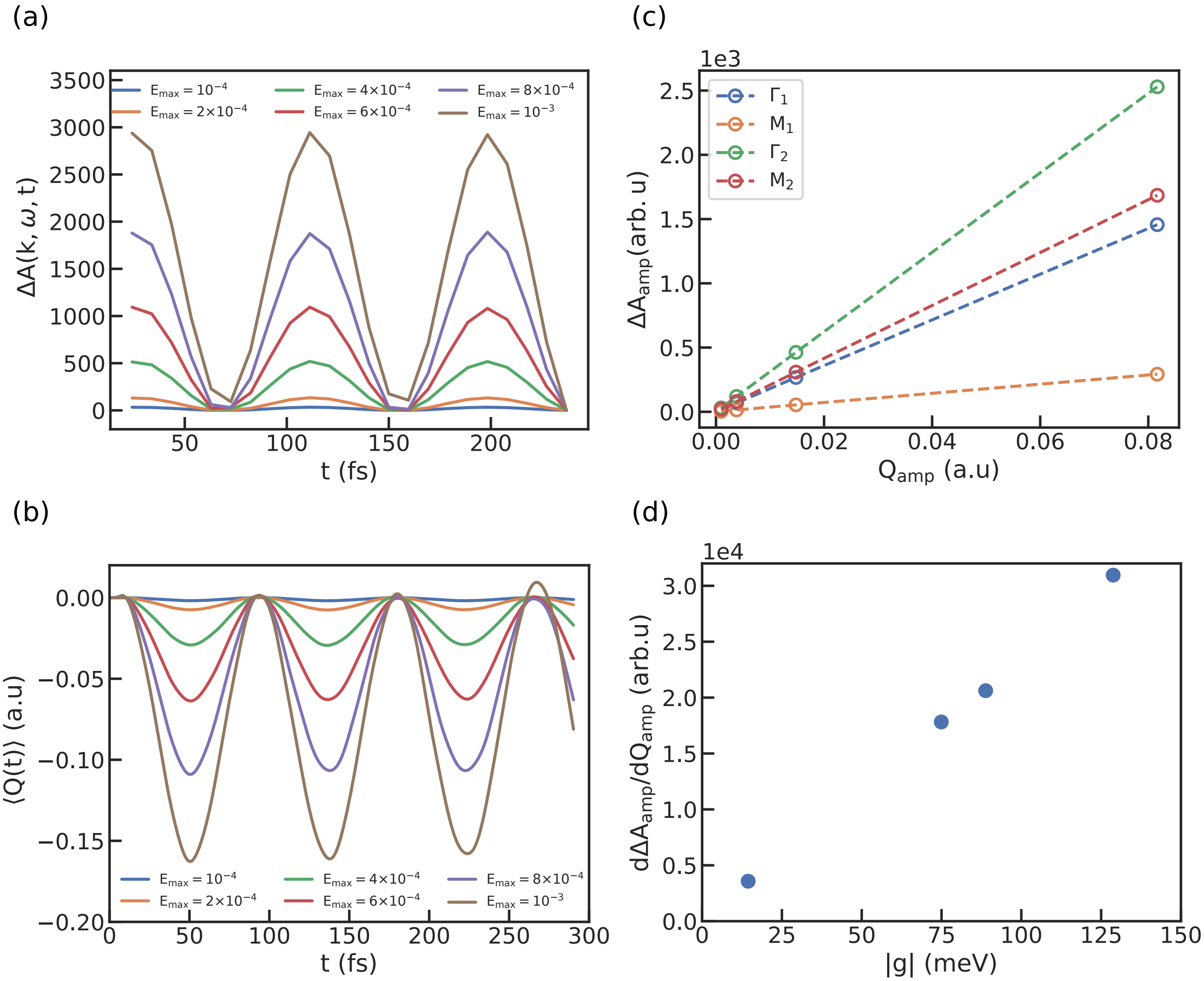

FIG. 2. (Color online) (a) Computed temporal evolution of spectrum intensity difference from the spectrum at $t$ = 237 fs evaluated at an energy of 1.70 eV at Γ at different pump intensity in the arbitrary unit. (b) Corresponding dynamics of the phonon displacement of the $A_1^{'}$ mode. (c) Dependence of $\Delta A_{amp}$ fitted from (a) on $Q_{amp}$ from (b) for selected valence states at high symmetry point. The subscript in the legend is the band index counted from Fermi level. Dashed lines are linear fitting results. (d) Dependence of the fitted slopes from (c) against e-ph coupling strength $g_{ii\nu}(\boldsymbol{k},\boldsymbol{q}=0)$ for $\nu = {A_1}'$ from first-principles calculations.

As a demonstration, we focus on monolayer $MoS_2$ in the H phase which exhibits strong excitonic effects. A large coherent-phonon-induced change in the optical absorption spectrum was observed around the exciton C peak in experiment [26]. At equilibrium, our GW-BSE calculations [29, 30] performed using the BerkeleyGW package [31] give an excitation energy for the lowest optically bright 'A' exciton of 1.76 eV and a band gap of 2.27 eV. Fig. 1 (a) shows the simulated TR-ARPES for monolayer $MoS_2$ with a 1.7 eV pump light (near resonant with the A exciton excitation energy) at a probe time of 237 fs after the pump pulse. We define $t = 0$ as the time when the pump starts as shown in Eq. S1 in SM [27]. Plotted is the spectral function $A(\boldsymbol{k}, \omega, t)$ with the piecewise discreteness due to a finite $\boldsymbol{k}$-point sampling in our calculations. The computed spectrum shows both the valence band dispersion and the exciton-Floquet features arising from the A exciton at energy an exciton excitation energy above the valence band maximum (VBM), as well as an exciton excited state in the same series [32]. We did not include the dissipation in our simulation to investigate the intrinsic dynamics. The effects of dissipation are demonstrated in the SM [27].

To identify the signatures of coherent phonons in TR-ARPES, we take the difference between the computed spectral function at an arbitrary time *t* and at *t* = 237 fs, $\Delta A(\boldsymbol{k},\omega,t) = A(\boldsymbol{k},\omega,t) - A(\boldsymbol{k},\omega,t=237\ fs)$. The results at the $\Gamma$ and $M$ points are shown in Fig. 1 (b) and (c), respectively. We can clearly see that the intensity of the difference spectrum (which has peaks and valleys near the equilibrium quasi-particle energies) modulates as a function of time, which indicates a periodic shifting of the quasiparticle state energy in time. A Fourier analysis of the intensity modulations shows that they have the same oscillation frequency as the $A_1'$ mode phonon frequency in monolayer $MoS_2$. The intensity of the various peaks/valleys in the difference spectrum $\Delta A(\boldsymbol{k},\omega,t)$ shows a strong electron state dependence. For example, in Fig. 1 (c) the difference spectrum intensity at energies near the highest valence band at $M$ is much weaker than that of the one corresponding to the second valence band at M (around -2.7 eV), which suggests coherent phonon motion only weakly couples to the highest valence band at $M$ as compared to the lower band. Furthermore, we observe that in Fig. 1 (b) the two states at $\Gamma$ (corresponding to the first and second valence band from the band gap) have energy shift in the opposite direction. In Fig. 1 (d) we show the temporal evolution of the light-pulse-induced phonon displacement expectation values of the three modes which we found to have non-negligible amplitudes. All three modes are Raman active, which is consistent with the known selection rule for optical coherent phonon generation. We find that, among these three modes, the $A_1'$ phonon mode has a much larger oscillation amplitude which is consistent having a single dominant modulation frequency from our Fourier analysis of the computed $\Delta A(\boldsymbol{k},\omega,t)$. Therefore, we will focus on the dynamics of the $A_1'$ mode in the following analysis.

Results from Fig. 1 suggest that we may extract information on the state-dependent e-ph coupling strength from the TR-ARPES difference spectrum. To further explore this possibility, we analyze the light field intensity dependence of coherent phonon amplitude and $\Delta A(\boldsymbol{k},\omega,t)$. In Fig. 2 (a) we show the energy cut of $\Delta A(\boldsymbol{k},\omega_0,t)$ taken at the energy $\omega_0$ where the intensity change is the largest near the first valence band at Γ, which clearly shows a periodic oscillation with a non- zero average value. To quantify the features in Fig. 2(a), we fit them to a function $\Delta A_{amp} sin(\omega_{ph} t + \phi_{\Delta A}) + \overline{\Delta A}$, where $\Delta A_{amp}$ is defined to be positive. The coherent phonon dynamics, that is, the phonon displacements $\langle Q_\nu(t)\rangle$, from the same set of calculations are shown in Fig. 2 (b). A similar function $Q_{\mathrm{amp}} sin(\omega_{ph} t + \phi_{\mathrm{Q}}) + \bar{Q}$ is also used to analyze the dynamics. Our fittings show that, as expected, both $\Delta A_{amp}$ and $Q_{amp}$ are proportional to the intensity of the applied pump field, $|E_{max}|^2$. Fig. 2 (c) shows the electronic state dependence of the change in the spectral shift $\Delta A_{amp}$ on the phonon displacement $Q_{amp}$ labeled by their high symmetry point symbol and a subscript of the valence band index counted from Fermi energy. We can clearly see the linear dependence between the two, which suggests the quasiparticle energy shift is proportional to coherent phonon amplitude. As all lines pass through the origin, the larger slopes correspond to larger $\Delta A_{amp}$. In Fig. 2 (d) we compare the fitted slopes against the e-ph coupling matrix elements $g_{nn\nu}(\boldsymbol{k},\boldsymbol{q}=0)$ of the corresponding state *n* for $\nu = A_1'$. The perfect proportionality confirms that $dA_{amp}/dQ_{amp}$ is proportional to $g_{nn\nu}(\boldsymbol{k},\boldsymbol{q}=0)$ and, in principle, the state-dependent e-ph coupling strength can be extracted from measuring $\Delta A(\boldsymbol{k},\omega,t)$ and Q. In passing, we note that the analysis could be performed for states at arbitrary **k** points although better spectral resolution is required if two bands are nearly degenerate. Complicated dynamics of states in resonant with the pump energy, e.g. those at $K$ and $K'$ in the current setup also pose a challenge to the fitting procedure.

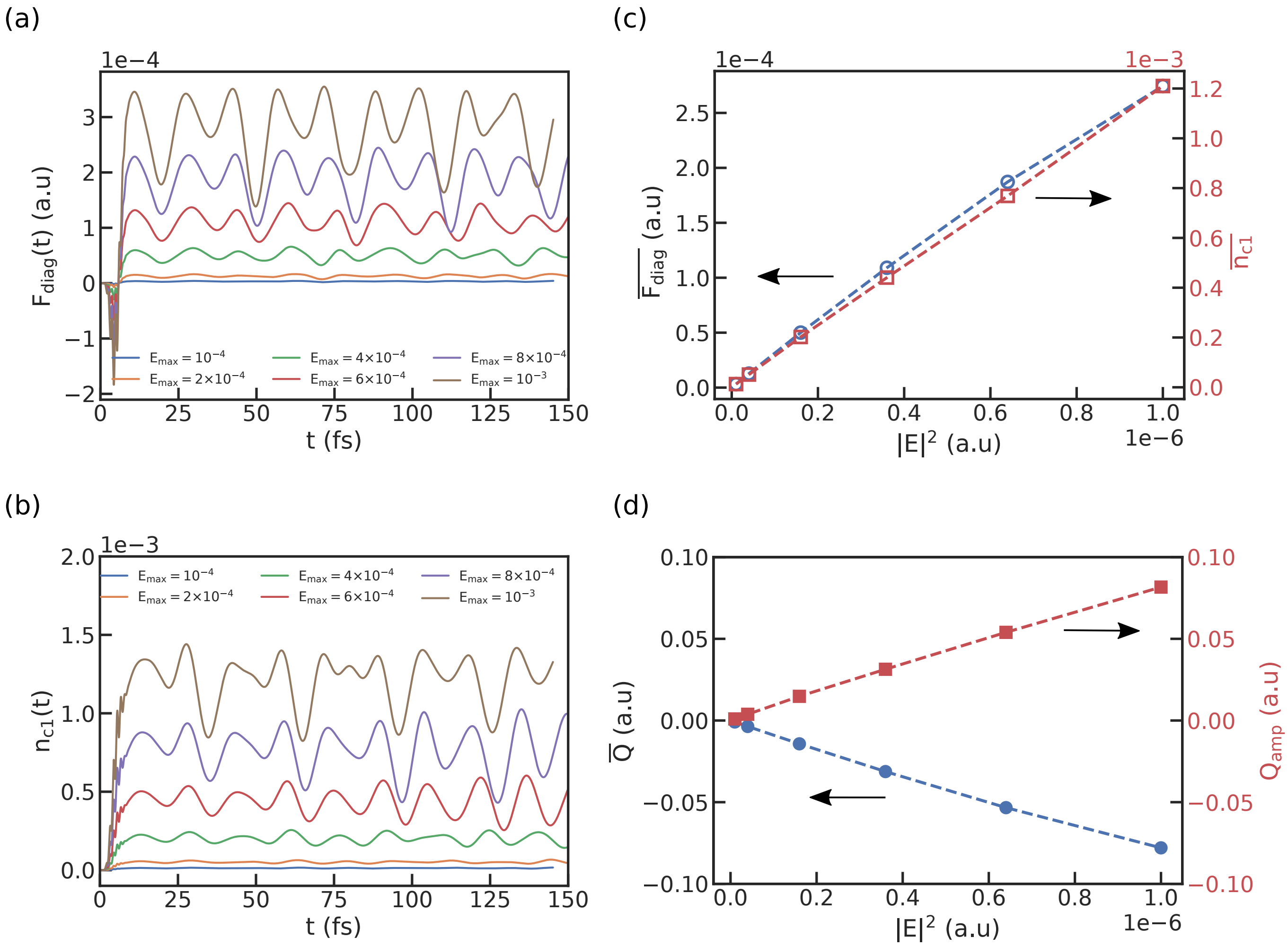

FIG. 3. (Color online) (a) The temporal evolution of the driving term $F_{diag}$ at different pump intensities. Pump frequency and duration are the same as those in Fig. 2. (b) Time evolution of the lowest conduction band electron density at the same setting as in (a). (c) Pump intensity dependence of the time averaged $F_{diag}$ (blue empty dots, left axis) and the lowest conduction electron densities (red empty squares, right axis) taken after 10 fs. (d) Dependence of $\bar{Q}$ (blue solid dots, left axis) and $Q_{amp}$ (red empty squares, right axis) on the pump intensity.

The finite values of $\bar{Q}$ and its proportionality to pump intensity hints that the DECP generation scenario, where excited carriers (as given in terms of the pump-pulse induced changes in the diagonal elements of the density matrix in the band orbital basis) exert a force and move the atoms to their new equilibrium position. To explore this mechanism, we focus on the sum of the diagonal part of the driving term (which we shall call the force) $F_{\nu,diag} \equiv -\frac{\omega_{0\nu}}{\hbar}\sum_{n\boldsymbol{k}} \tilde{g}_{nn\boldsymbol{k}}\rho_{nn\boldsymbol{k}}$ for the $\nu = A_1{}'$ mode at different pump intensities as depicted in Fig. 3 (a) since we numerically find that the off-diagonal terms play a secondary role in coherent phonon dynamics as fast oscillations cancel their contributions. We observe two features in the temporal behavior of $F_{diag}$: (1) A dip during the first 10 fs after the arrival of the pump pulse (pump starts at $t = 0$), with magnitude closely follows the intensity of the pump field, and (2) oscillations around a finite average value afterward. The latter leads to the finite $\bar{Q}$ observed in Fig. 2 (b). The finite average force can be further connected, for our present case of a pump frequency of 1.7 eV, to the excitation of excitons which leads to a finite value in the diagonal elements of the density matrix in the conduction states (i.e., finite photoexcited carrier density) shown in Fig. 3 (b) where we plot the temporal dependence of the sum of the density matrix element ($\sum_{\boldsymbol{k}} \rho_{nn\boldsymbol{k}}$) of the first conduction band at different pump intensities.

We further show the pump intensity dependence of the averaged force, conduction band electron density, $Q$, and $Q_{amp}$

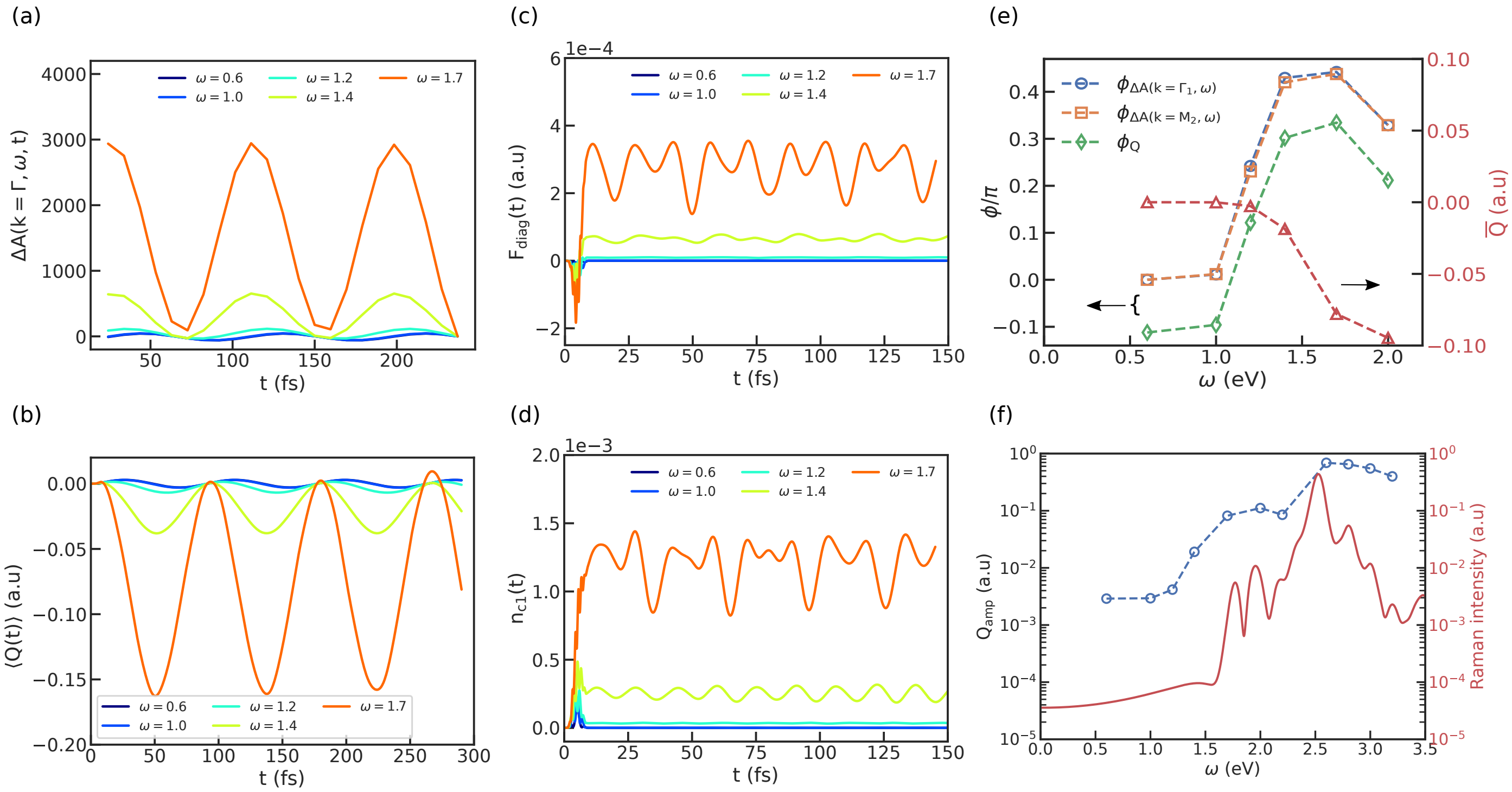

FIG. 4. (Color online) The temporal dependence of (a) the TR-ARPES spectrum intensity difference $\Delta A(\boldsymbol{k},\omega,t)$ for the first valence band at $\Gamma$, (b) coherent phonon displacement of ${A_1}'$ mode, (c) $F_{diag}$, and (d) the lowest conduction band electron densities, computed at different pump frequencies. Frequency indicated is in the unit of eV. Calculations were done with parameters the same as those for Fig. 1-3, but with $E_{max} = 10^{-3}$ a.u. (e) Left axis: Pump frequency dependence of the fitted phase angle $\phi$ of $\Delta A(\boldsymbol{k},\omega,t)$ of the first valence band at $\Gamma$ (blue empty dots), of $\Delta A(\boldsymbol{k},\omega,t)$ of the second valence band at $M$ (orange empty squares), and of the coherent phonon displacement $\phi_Q$ (green empty diamonds); Right axis: Pump frequency dependence of the mean displacement $\bar{Q}$ (red empty triangles). (f) Amplitude of coherent phonon generated (blue empty dots, left axis) as a function of pump frequency and Raman intensity (red solid line, right axis) as a function of light frequency.

after the pump pulse in Fig. 3 (c) and (d). The proportionality of all these quantities to pump intensity can be well-accounted for from the EOM of the phonon displacement, Eq. 4. First, the temporal profile of the force follows that of the density matrix $\rho_{nm\boldsymbol{k}}$. For the ${A_1}'$ mode, we find that the e-ph matrix elements $g_{nm\nu}(\boldsymbol{k},\boldsymbol{0})$ are dominantly diagonal in band indices for states close to band edges at $K$ and $K'$ in monolayer $MoS_2$, which are the states that can form electron-hole pairs with transition energy (≥2.27 eV) close to the pump energy 1.7 eV. Therefore, the temporal dynamics of the diagonal force follows that of the change in the diagonal elements of the density matrix which we loosely call it the band electron density. Second, it can be shown that the excited electron density is proportional to the pump intensity from light absorption [27], which further explains why the diagonal force is also proportional to the pump intensity. The linear dependence between the $\bar{Q}$ and the mean value of the driving force can be seen by taking the time-average over both sides of Eq. 4. From these results we conclude that the coherent phonon generated with a 1.7 eV pump frequency is due to the e-ph coupling from the excited carriers, which is in line with the DECP mechanism [18].

Next, we investigate the pump frequency dependence of the TR-ARPES difference spectrum and coherent phonon. Fig. 4 (a) and (b) show the temporal evolution of $\Delta A(\boldsymbol{k},\omega,t)$ for the first valence band at $\Gamma$ and that of $\langle Q(t)\rangle$ as the pump frequency varies, respectively. For a pump frequency of 1.4 eV, we observe results similar to those from a 1.7 eV pump frequency although with a smaller $\Delta A_{amp}$ and $Q_{amp}$. The corresponding time-dependent force and carrier density are shown in Fig. 4 (c) and (d), respectively. In particular, the finite value of the averaged force and carrier density after the pass of the pump pulse are consistent with a DECP generation mechanism. On the other hand, the $\Delta A(\boldsymbol{k},\omega,t)$ and $\langle Q(t)\rangle$ of the same state from pump pulse with frequencies of 0.6 eV and 1.0 eV clearly show distinct features. Besides the significantly smaller oscillation amplitude, we find that the oscillations are around an average value of zero as shown in Fig. 4 (a) and (b), respectively. The pump frequency dependence of the extracted phase shift $\phi_{\Delta A}$ for the first state at $\Gamma$ and the second state at $M$, as well as $\phi_Q$, and $\bar{Q}$ are shown in Fig. 4 (e). We note that the exact value of $\phi_{\Delta A}$ depends on the reference probe delay time at which we take the spectrum difference. Therefore, we plot $\phi_{\Delta A}$ relative to the value from that of pump frequency 0.6 eV. In contrast, the value of $\phi_Q$ is well-defined in the sense that we always start the simulation with $\langle Q(t=0)\rangle = 0$. Figure 4 (e) clearly reveals two different regions depending on whether the pump frequency is far below or close to the optical gap. At low pump frequency, both the $\phi$ values and the time averaged phonon displacement $\bar{Q}$ are close to zero. When the pump frequency increases, we see a change of $\Delta\phi = \pi/2$ in phase shift relative to their low frequency value along with acquiring a finite value for $\bar{Q}$. The distinct characters at low pump frequency suggest a coherent phonon generation mechanism different from DECP. In the SM [27], we further include the results from a pump pulse with 20 fs duration. Although the exact change over frequency depends on the pulse duration, we can clearly identify two different regions with characters similar to those observed here.

To investigate the generation mechanism at low frequency, we show the time-dependent force and conduction band electron density in Fig. 4 (c) and (d), respectively. We observe that for pump frequency below 1.0 eV the diagonal force drops to zero immediately after the passing of the pump pulse. At the same time there is no carrier density left after the pump pulse has passed, which contrasts with the cases with frequencies closer to resonant with the exciton excitation energy. In the low frequency region, the driving force and density simply follow that of the pump intensity with little oscillations. These features are consistent with the ISRS mechanism, which is known to apply to materials that are transparent to the pump pulse. However, we find that there is no clear-cut transition frequency between the two regions of frequencies, as we observe small, but finite carrier density builds up after the pump pulse and a phase shift between zero and $\pi/2$ near an intermediate pump frequency of 1.2 eV. We note that with an above band gap pump frequency, our calculations show that a simple fitting to a sinusoidal function for $\Delta A(\boldsymbol{k},\omega,t)$ and $\langle Q(t)\rangle$ is challenging as the amplitude become time dependent which could be due to the complicated electron dynamics (see SM [27]).

Observations of features of both DECP and ISRS mechanism for coherent phonon generation in our simulations (depending on pump frequencies) and intermediate behaviors hint that the two mechanisms are not mutually exclusive. From Eq. 4, we can see the response of the density matrix to the pump light governs the driving force and the coherent phonon dynamics to first order in the e-ph couplings, from which we can deduce the selection rule. In particular, from second order perturbation theory, the light-excited carrier density is proportional to the pump intensity and the matrix product of two optical matrices [27,33]. Therefore, having a non-zero driving force is equivalent to having a nonzero matrix product $\sum_{nml} g_{nm\nu}(\boldsymbol{k},0)\,\boldsymbol{d}_{ml\boldsymbol{k}}\boldsymbol{d}_{ln\boldsymbol{k}}$, where $\boldsymbol{d}_{nm\boldsymbol{k}}$ is an optical matrix element. Coincidently, the same matrix product also appears in the numerator of the expression for the Raman scattering tensor [27,34-37]. Therefore, photogeneration of coherent phonon is generally observed for phonon modes that are Raman active. However, the two spectra in general are not the same, which is evident from the full expression of the Raman scattering tensor and the driving force term. A perturbative expression of the driving force is given in the SM [27], which shows the different resonance structure when compared against those of the Raman scattering intensity. As a demonstration, we show in Fig. 4 (f) the frequency dependence of the coherent phonon displacement amplitude (excited at a pump intensity of $10^{-3}$ a.u.) and the computed Raman scattering intensity spectrum. Although both spectra show secondary peaks at energy in resonant with the A and B excitons and a largest peak at the C exciton energy, which agrees with previous experiment and calculations [26], the two spectra deviate at frequency below the optical gap and above the C peak energy.

In our simulations, we assume coherent dynamics for both the electron and phonon degree of freedom, which simplifies the discussion and allows us to clearly identify the generation mechanism and the effects of coherent phonons on the TR-ARPES. In practice, dephasing and depopulation of excitations can be relevant in the hundred-fs time scale. In our study, we have also conducted some calculations including finite dephasing and depopulation rates in the order of 0.01 to 0.05 eV [27]. In such simulations, we find that the modulations in both the coherent phonon displacement and the TR-ARPES spectrum have zero average value following the fast decay of excited carriers. At the same time, the fitting of $\phi$ value is more involved and ambiguous in identifying the generation mechanism in experiments. Also, here, we have focused on the coherent phonon generation from the density response at second order of electron-light couplings. Responses from other orders of light-matter couplings and higher order e-ph or phonon-phonon couplings are also possible [22], which are interesting future directions of study.

## IV. CONCLUSION

In conclusion, we have demonstrated from our first-principles time-dependent calculations including both e-h and e-ph couplings that coherent phonon generation and state-resolved e-ph couplings can be revealed and analyzed through coherent-phonon modulations in pump-probe TR-ARPES experiments. We demonstrate the procedure with monolayer $MoS_2$ as an example and identify signatures that are characteristic of both ISRS and DECP coherent phonon generation, depending on the optical pump pulse frequency. We show that the two mechanisms originate from the same density matrix response in two frequency regions. We further provide an understanding of the coincident selection rule of Raman scattering and coherent phonon generation.

## ACKNOWLEDGEMENT

Y.H.C. thanks Cheng-Tien Chiang and Fabio Caruso for their helpful discussion. Y.H.C. is supported by the National Science and Technology Council of Taiwan under grant no. 110-2124-M-002-012 and the Academia Sinica Career Development Award under project no. AS-CDA-114-M04. This work is supported by the Center for Computational Study of Excited-State Phenomena in Energy Materials (C2SEPEM) at the Lawrence Berkeley National Laboratory (LBNL), which is funded by the U.S. Department of Energy (DOE), Office of Science, Basic Energy Sciences, Materials Sciences and Engineering Division under Contract No. DEAC02-05CH11231, as part of the Computational Materials Sciences Program, which provided the development of methods and advanced codes, and is also partially supported by the National Science Foundation under Grant No. OAC—2513830 in the development of interoperable software enabling the EPW and BerkeleyGW calculations with consistent gauge. Z.L. acknowledges the support from the U.S. National Science Foundation (NSF) CAREER Award under Grant No. DMR-2440763. We acknowledge the use of computational resources at National Center for High-performance Computing (NCHC).